# Dependence of the thermoluminescent high-temperature ratio (HTR) of LiF:Mg,Ti detectors on proton energy and dose

P. Bilski[1], M. Sadel[1], J. Swakon[1], A. Weber[2]

[1] Institute of Nuclear Physics, Radzikowskiego 152, 31-342 Krakow, Poland
[2] Charite-Universtatsmedizin Berlin, BerlinProtonen am Helmholtz-Zentrum Berlin, Hahn-Meitner-Platz 1,14109 Berlin, Germany

**Abstract**

The high-temperature ratio (HTR) is a parameter quantifying changes of the shape of the high-temperature part of the LiF:Mg,Ti glow-curve after exposure to densely ionizing radiation. It was introduced in order to estimate the 'effective LET' of an unknown radiation field and to correct the decreased relative TL efficiency for high-Linear Energy Transfer (LET) radiation.

In the present work the dependence of HTR on proton energy (14.5 - 58 MeV) and dose (0.5 – 30 Gy) was investigated. All measured HTR values were at the level of 1.2 or higher, therefore significantly different from the respective value for gamma-rays (HTR=1), but HTR was found to be insensitive to changes of proton energy above 20 MeV. As a result the relationship between HTR and relative TL efficiency is not unequivocal. The HTR was found to be dependent on absorbed dose even for the lowest studied doses.

Keywords: Thermoluminescence, High-temperature emission, HTR, LiF:Mg,Ti

## 1. Introduction

The high-temeprature ratio (HTR) is a parameter quantifying changes in the shape of the high temperature part of the LiF:Mg,Ti glove curve after exposure to densely ionizing radiation. It is defined as the ratio of thermoluminescent signal integrated over a defined temperature range after exposure to studied radiation and after reference gamma exposure (see Eq. (1), and Fig. 1.).

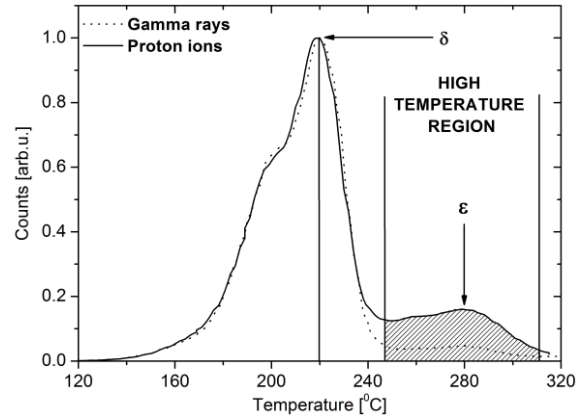

**Fig. 1.** The definition of HTR. Symbol δ represents a maximum height of the main dosimetric peak and ε is an integral of the high-temperature region (see Eq. 1.)

$$HTR = \frac{\varepsilon_k/\delta_k}{\varepsilon_\gamma/\delta_\gamma} \qquad (1)$$

The HTR method is based on the assumption that unequivocal functional relationships exist between the HTR parameter and LET and between HTR and relative TL efficiency. These relationships may be used to estimate LET in an unknown radiation fields (Vana et al., 1996), as well as to correct the decreased relative TL efficiency for high-LET radiation (Berger et al., 2006a). The limitations of the HTR method were recently discussed (Bilski, 2010). It was demonstrated that in the general case of a mixed radiation field, the estimation of LET is unreliable (however, it may be correct for simple radiation fields, like a single type particle beams). On the other hand the correction of the relative TL efficiency with the HTR produces quite good results. It should be also mentioned that this method is purely empirical with no significant theoretical argumentation. The drawbacks of the method are lack of universality of the HTR characteristic and non-linearity of the HTR dose – response relationship (Horowitz et al., 2003,2007).

The HTR was several times measured and reported for heavy charged particles, ranging from helium up to xenon ions. There was however not much attention paid to HTR applied for protons. Only Schöner et al., (1999) presented HTR data for 62 MeV and 10 MeV protons. This is probably due to the fact, that increase of the high-temperature peaks after exposure to protons, while noticeable, is not very pronounced. On the other hand application of the HTR method to proton dosimetry could be quite important, as proton radiotherapy becomes more and more widely used technique. TLDs are frequently applied for in-phantom measurements of proton doses, and a change of the relative TL efficiency due to a change of proton energy may increase dose measurement uncertainty

Therefore, the goal of the present work was to investigate the dependence of HTR on proton dose and energy. The experiments were carried out at the Institute of Nuclear Physics (IFJ) in Krakow and at the Helmholtz Centrum Berlin, exploiting the mono-energetic proton beams normally used for eye tumour therapy.

## 2. Materials and methods

### 2.1. Proton irradiations

The irradiations were realized at the Proton Eye Radiotherapy Facility at the IFJ with 60 MeV proton beam (Swakon et al., 2010) and at the Helmholtz Centrum Berlin with 72 MeV proton beam (Denker et al., 2010). A uniform lateral dose distribution was achieved by passive scattering with a single tantalum foil. The proton beam range was controlled with a PMMA range shifter. During irradiation proton dosimetry was carried out with Markus ionization chambers coupled with PMMA moderators of variable thickness or moved inside water phantom using 3D scanner. Measurements of the dose-depth distribution were performed with a resolution better than 0.1 mm. The irradiations were performed from 59 MeV down to energy of 14.5 MeV (corresponding to a proton range of 1.19 mm in 2.5 g.cm$^{-3}$ LiF). For the assessment of dose absorbed in the dosimeters were obtained with a Markus ionization chamber. The proton dose was at the level of 0.5 Gy in all cases, to avoid effects of supralinearity. The dose rate was about 0.1 Gy/s.

In a separate experiment at the Institute of Nuclear Physics the dependence of the linearity index $f(D)$ on proton dose was investigated. In this measurement for energy around 16 - 17 MeV (corresponding to a proton range of around 1.50 mm in 2.5 g.cm$^{-3}$ LiF) TLDs were irradiated with a dose range from 0.5 Gy to 30 Gy.

The energy and LET of protons for a given depth in water were calculated in water using the SRIM code (Ziegler et al., 2010).

### 2.2. TL detectors

LiF:Mg,Ti (MTS-N) thermoluminescent detectors in form of sintered pellets with dimensions 4.5x0.9 mm and density 2.5 g.cm$^{-3}$ manufactured at the IFJ Krakow were used. The following annealing conditions were applied 400$^{o}$C/1h + 100$^{o}$C/2h. The readout system was a Harshaw Series 3500. Before readout the detectors were preheated at a temperature of 100$^{o}$C for 10 min. TL glow-curves were registered by heating up to 350 $^{o}$C at a heating rate of 5 $^{o}$C.s-1. Calibrations of TLDs (conversion of TL signal to gamma-ray dose) were performed by irradiating a group of TLDs from each batch with a dose of $^{137}$Cs gamma rays at the secondary standard calibration laboratory at IFJ and using Co-60 gamma rays in case of dose response measurements. Additionally, to minimize and to correct any spread of sensitivity between TLDs pellets of one group, individual response factors for each detector were determined. The HTR part of TL glove curves was calculated for the temperature range between 248 $^{o}$C - 310 $^{o}$C according to approach of Berger at al., (2006a), while relative TL efficiency was evaluated basing on the integral of the main peak 5 (from 100$^{o}$C to 240$^{o}$C with main peak position at 220$^{o}$C).

## 3. Experimental results and discussion

For the evaluation of the HTR and the relative TL efficiency a group of dosimeters was irradiated in the proton beams with energy ranging from 14.5 to 58 MeV. In this region of energy, the whole volume of detectors is irradiated. The results are summarized in Figure 2. The vertical error bars give the statistical error from measurements where more dosimeters were exposed under the same condition.

Figure 2a and figure 2b, show the HTR ratio versus proton energy and LET respectively. For low LET values, up to about 2.5 keV.um$^{-1}$, which correspond with the proton energy exceeding 20 MeV, the HTR values is approximately constant at the level of 1.2. For higher LET HTR increases reaching 1.55.

The obtained values of HTR are in general agreement with the result of Berger and Hajek (2008) and Bilski (2010). It is somewhat surprising that HTR seems to be not dependent on LET in low-LET region, i.e. for higher proton energies, while difference between protons and gamma-rays (HTR=1 by definition) is quite significant.

Measurement presented in Fig. 2c shows that the relative TL efficiency is approximately constant at the level 1.09-1.12 for proton energies above 30 MeV. Similar data were presented by Sądel et al., (2013). For lower energies the efficiency increases, reaching a maximum of 1.18-1.20 for about 18-14 MeV.

Figure 2d illustrates relationship between HTR and relative TL efficiency. As was mentioned, this relationship should be unequivocal in order to use it for correcting TL efficiency. The obtained results indicate that for low HTR values (below c.a. 1.3), such correcting is not possible.

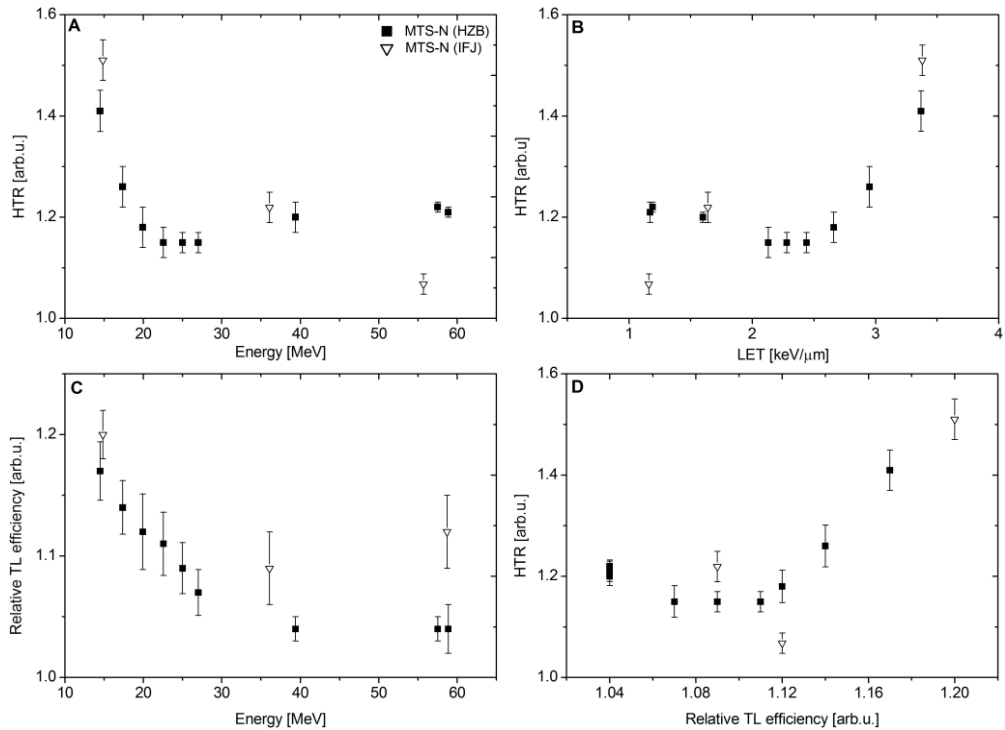

**Fig. 2 .** Data calculated for MTS-N detectors. Panel A and panel B shown comparison of the dependence of the HTR ratio on energy and LET respectively. Within panel C,

dependence of the relative TL efficiency on energy was presented. On the panel D, the relationship between the HTR and the relative TL efficiency is presented. Corresponding open symbols present the result from irradiation at IFJ. Full symbols present results from HZB irradiations.

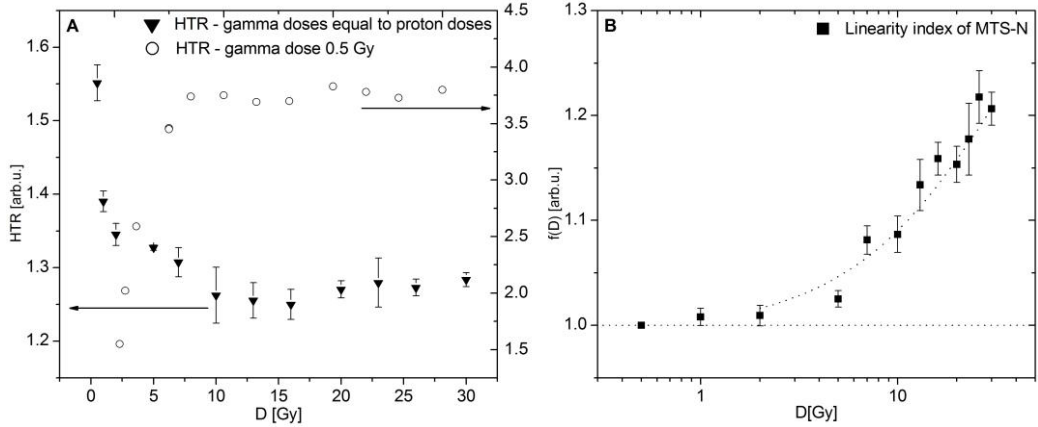

**Fig. 3.** HTR for LiF:Mg,Ti detectors (panel A), and linearity index (panel B) obtained for 17 MeV proton irradiations for dose range at the level of 0.5 to 30 Gy. Irradiations were performed at the IFJ PAN. Dashed line in panel B indicates the linear trend.

It is well known that LiF:Mg,Ti shows supralinear dose response above about 1 Gy. It is also known that onset of supralinearity of the high-temperature peaks for gamma-rays is about 100 mGy. To study influence of proton dose on HTR, irradiations with 17 MeV proton beam were realized in the range from 0.5 to 30 Gy (lower doses were not available due to technical limitations of the accelerator). Simultaneously TLDs were also exposed to identical doses of Co-60 gamma-rays. The HTR values were calculated in two ways: using the gamma-ray data for the same dose as the respective proton dose (method A) and using the dose 0.5 Gy, which was normaly applied for calibration (method B). The results are presented in Figure 3a, while figure 3b presents data on supralinerity of the main dosimetric peak in form of linearity index $f(D)$ calculated according to the following formula:

$$f(D) = \frac{I(D)/D}{I(D_0)/D_0} \qquad (2)$$

where $I$ is the intensity of TL signal and $D_0$ is the dose from the linear dose response range. Error bars represent standard deviations between the results obtained for each irradiated detector for each point in measured dose range. It is clearly visible that the HTR ratio calculated with the method A steeply decreases with increasing dose, from about 1.55 for 0.5 Gy down to 1.2-1.3 for 10 Gy. For higher doses HTR remains on approximately the same level. This effect is a result of faster supralinear growth of high-temperature peaks for gamma-rays, than for protons. The HTR values calculated with the method B, behave oppositely: there is a very steep increase up to 6 Gy and for higher doses HTR remains constant. The observed nonlinear behavior of HTR raises doubts about possibility of applying of HTR in a mixed radiation field where various doses of protons with various energies may be encountered.

## 4. Conclusions

The HTR ratio of LiF:Mg,Ti TL detectors was determined for protons with energy ranging from 14.5 to 58 MeV and for doses ranging from 0.5 Gy to 30 Gy. All measured HTR values were at the level of 1.2 or higher, therefore significantly different from the respective value for gamma-rays (HTR=1). On the other hand changes of HTR with proton energy were noticeable only for low energies, below 20 MeV. As a result the relationship between HTR and relative TL efficiency is not unequivocal.

HTR was found to be dependent on absorbed dose even for the lowest studied doses. When gamma-ray calibration was done with doses equal to proton doses, this dependence was changed, but still remained nonlinear.

The lack of unequivocal relationship between HTR and TL efficiency, and nonlinear dose characteristic, make possibility of application of the HTR method therapeutic proton beams rather difficult.

**Acknowledgments**

This work was supported by the National Science Center (project DEC-2011/01/B/ST2/02450)